# High-Order Nonreciprocal Add-Drop Filter

Hang Li, Rui Ge, Yuchen Peng, Bei Yan, Jianlan Xie, Jianjun Liu*, and Shuangchun Wen

*Key Laboratory for Micro/Nano Optoelectronic Devices of Ministry of Education & Hunan Provincial Key Laboratory of Low-Dimensional Structural Physics and Devices, School of Physics and Electronics, Hunan University, Changsha 410082, China*

Topological photonics have led to the robust optical behavior of the device, which has solved the problem of the influence of manufacturing defects and perturbations on the device performance. Meanwhile, temporal coupled-mode theory (t-CMT) has been developed and applied widely. However, the t-CMT of cascaded coupling cavities (CCC) system and its corresponding high-order filter has yet to be established. Here the t-CMT of CCC system is established based on the existing t-CMT. By combining the CCC with the topological waveguides, a versatile design scheme of the high-order nonreciprocal add-drop filter (HONAF) is proposed. The relationship between coupling effect of cavities and transmission and filtering performance of HONAF is analyzed quantitatively, then a method to improve the trans-mission efficiency and quality factor of the filter is given. Based on the combination of gyromagnetic photonic crystals and decagonal Penrose-type photonic quasicrystals, a HONAF is proposed. The transmission and filtering performance of the HONAF are numerically analyzed, which verifies the consistency between theoretical prediction and numerical simulation. The t-CMT of CCC system established can be widely used in coupled resonator optical waveguides and their related systems. The HONAF proposed with excellent performance can also be applied and compatible to wavelength division multiplexing/demultiplexing system.

## I. INTRODUCTION

Wavelength division multiplexing/demultiplexing (WDM) system and optical integrated circuits with large density, high speed transmission and low energy consumption have developed rapidly [1-2]. With the continuous development of micronano lithography and subwavelength structure manufacturing technologies [3-4], novel micronano photonics devices have been proposed, such as fibers [5-8], sensors [6], lenses [9,10], couplers [11], absorbers [12-15], resonators [16-21], and other photonic crystal (PC) devices [22]. Optical resonator is the core part of many micronano photonics devices due to its excellent natural frequency selection characteristics, such as coupled resonator optical waveguides (CROWs) [16], filters [23], sensors [24], lasers [25], etc. The frequency-division/recombination of signals in optical filters plays an irreplaceable role in dense WDM and optical integrated circuits. Therefore, the design and characteristics of optical filters have been researched for a long time [18,23,26-32].

At present, the filters constructed by a single cavity have the characteristics of multiple drop channels, nonreciprocity, high transmission efficiency ($T$) and quality factor ($Q$), and the radiation loss inherent in the PC cavities has been reduced to a low level [17-20,26,32]. However, these filters are unable to immune to the defects in waveguides or cavities, and the transmission efficiency is sensitive to the coupling phase shift between waveguides and cavities, which brings great troubles to device fabrication and application [16,20,27]. Furthermore, the performance of the single cavity filters for the out-of-band rejection ratios and flat tops bandwidth tunable property is poor. In fact, the practical function also needs to be further expanded [17,18,20,26,27]. In order to overcome these problems, a topological nontrivial waveguide and cavity was introduced into PC filters, which can be immune to the defects in structure, bringing the transmission characteristic of reciprocity, and the problem that the transmission efficiency is sensitive to the coupling phase shift of optical waveguides and cavities is solved [19,26]. Therefore, PCs with topological properties are widely used, such as isolators [33], one-way waveguides [34-46], resonators [19,21,47-49], lasers [50], logic gates [51,52,53], etc. In addition, the structure of the filters cascaded with microring resonators can improve the out-of-band rejection ratios, possess flat tops bandwidth tunable property and realize multiple filtering, but its structure is more complex and the composition of the material is not common [28,29]. Moreover, the existing temporal coupled-mode theory (t-CMT) can only accurately predict the transmission efficiency and quality factor of a single cavity or a small number of cavities coupling systems [22,26]. In order to make it more universal, that is, to predict the transmission efficiency and quality factor of any cascaded coupling cavities (CCC) system, and to improve its structure to optimize performance parameters, it is indispensable to explore and derive the applicative t-CMT equations for infinite CCC system.

---

*Corresponding author.  jianjun.liu@hnu.edu.cn

In Section II, based on the existing t-CMT, the t-CMT of CC system is derived. The transmission and filtering characteristics of four-channel high-order nonreciprocal add-drop filter (HONAF) coupled by topological PC waveguides and the CCC are deeply researched to find a multifunctional and high-performance filter and a universal design scheme. In Section III, in order to verify the strictness and practicability of theoretical analysis, combining the gyromagnetic photonic crystals (GPCs) [34,35,41] and photonic quasicrystals (PQCs) [54,55] (excellent frequency selectivity) with abundant defect mode properties, a composite HONAF working at microwave frequency band is proposed, whose performance index is analyzed when the number of cascaded cavities varies. Point defects (size and material of several scatterers changed) and line defects are introduced to verify the properties of the waveguide channels protected by topology. The main work of this paper is briefly summarized in the Section IV.

## II. t-CMT OF CCC SYSTEM

Weak coupling is an indispensable condition of the t-CMT, which guarantees the evanescent wave coupling relationship between waveguides and cavities, and the favourable frequency selection behavior of cavities. Weak coupling can be achieved by placing enough scatterers around the cavities [22]. Therefore, it is assumed that the relationship between waveguide channels and the cascaded cavities in the HONAF is weakly coupled. The operating principle of the HONAF system is shown in Fig.1. Two non-reciprocal waveguides constitute the bus waveguide $W_1$ and the add/drop waveguide $W_2$ respectively, which is coupled with cascaded cavities to form a four-port system. The signal in the bus waveguide passes through the frequency selection behavior of the cavities and drops the signal of the specific frequency to the $P_3$ or $P_4$ port determined by the nonreciprocal direction, i.e. the signal drop channel of the HONAF system can be transformed by nonreciprocal direction. Moreover, if the light wave signal is input from the add/drop waveguide channel, the function of add signal can be realized.

For convenience, two nonreciprocal waveguides

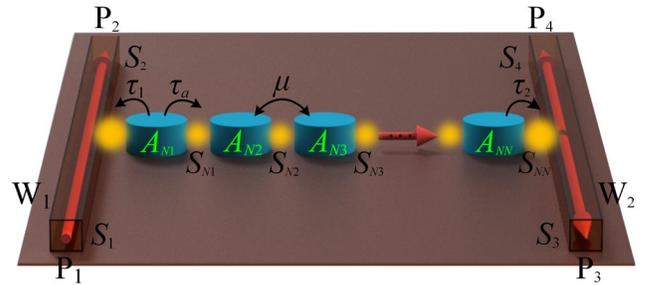

FIG. 1. Schematic diagram of the HONAF. $W_1$ and $W_2$ represent bus and add/drop waveguides, respectively; $P_1$, $P_2$, $P_3$ and $P_4$ represent the four ports of the HONAF, and the blue discs represent the cavities; $S_1$, $S_2$, $S_3$ and $S_4$ represent normalized amplitudes of the input and output fields, respectively; $S_{Ni}$ ($i=1,2,…,N$) represents normalized amplitudes of the cavity output fields; $A_{Ni}$ represents normalized amplitudes of the cavity fields; $\mu$ represents the interaction factor between the nearest-neighbor cavities; $\tau_1$, $\tau_2$ and $\tau_a$ represent the lifetimes of the cavity fields decay into the bus waveguide, the add/drop waveguide and the nearest-neighbor cavities, respectively.

and cascaded cavities are both structurally and functionally symmetric, which supports only one mode within the desired frequency range. When the field source is placed on port P$_1$ and the drop port is set as P$_4$ [Fig.1], then the transmission efficiencies of all the ports of the HONAF (derivation process see Section I of the Supplemental Material [56]) are

$$T_{2(1)} = \left|\frac{S_2}{S_1}\right|^2 = \left|1 - \frac{2/(\tau_1\tau_2)^{1/2}}{j2\pi(f-f_0)+1/\tau_0+1/\tau_1+1/\tau_2}\right|^2, \quad (1)$$

$$T_{2(N)} = \left|\frac{S_2}{S_1}\right|^2 = \left|1 - \frac{2/(\tau_1\tau_2)^{1/2}(-j\mu+2/\tau_a)^{N-1}}{dH_{N-1}+cF_{N-1}}\right|^2 \quad (N \geq 2), \quad (2)$$

$$T_{3(N)} = \left|\frac{S_3}{S_1}\right|^2 = 0 \quad (N \geq 1), \quad (3)$$

$$T_{4(1)} = \left|\frac{S_4}{S_1}\right|^2 = \left|\frac{2/(\tau_1\tau_2)^{1/2}}{j2\pi(f-f_0)+1/\tau_0+1/\tau_1+1/\tau_2}\right|^2, \quad (4)$$

$$T_{4(N)} = \left|\frac{S_4}{S_1}\right|^2 = \left|\frac{2/(\tau_1\tau_2)^{1/2}(-j\mu+2/\tau_a)^{N-1}}{dH_{N-1}+cF_{N-1}}\right|^2 \quad (N \geq 2), \quad (5)$$

where,

$$dH_{N-1}+cF_{N-1} = \frac{abd+2ac+2cd-bc+(ad+c)(b^2+4c)^{1/2}}{2^{N-1}(b^2+4c)^{1/2}\left[b+(b^2+4c)^{1/2}\right]^{2-N}}$$
$$-\frac{abd+2ac+2cd-bc-(ad+c)(b^2+4c)^{1/2}}{2^{N-1}(b^2+4c)^{1/2}\left[b-(b^2+4c)^{1/2}\right]^{2-N}}, \quad (6)$$

$$a = j2\pi(f-f_0)+1/\tau_0+1/\tau_1+1/\tau_a, \quad (7)$$
$$b = j2\pi(f-f_0)+1/\tau_0+2/\tau_a, \quad (8)$$
$$c = j\mu(-j\mu+2/\tau_a), \quad (9)$$
$$d = j2\pi(f-f_0)+1/\tau_0+1/\tau_2+1/\tau_a, \quad (10)$$

where, the physical significances of parameters $a$, $b$, $c$ and $d$ are complex frequencies. $j$ is an imaginary unit, and $f$ is the mixed frequency of the input time harmonic field. $f_0$ is the resonance frequency of a single isolated cavity, and $N$ represents the number of cascaded cavities (i.e. the order of the HONAF). According to the definitions of $\tau_a$ and $\mu$, the change in $\mu$ will lead to the change in $\tau_a$, assuming that

$$\tau_a = \hat{A}\mu, \quad (11)$$

where $\hat{A}$ is the operator factor, representing the relationship between $\tau_a$ and $\mu$. Generally, $\tau_a$ increases with the increase of $\mu$. The specific functional relationship between them needs to be obtained according to the specific structure.

Then, the transmission efficiency of drop port P$_4$ is specifically analyzed, i.e. Eq. (5). If $a$, $b$, $c$ and $d$ are substituted into Eq. (5), the higher-order function of "$f$-$f_0$" can be obtained, and the highest-order is $N$. Specifically, the numerator of the $T_{4(N)}$ expression is a constant, while the denominator is a high-order polynomial in terms of "$f$-$f_0$". Therefore, within a large frequency range, the $T_{4(N)}$ transmission spectrums will possess multiple formants. According to the definition of transmission efficiency of filter [22], $T_{\max}=1$. Therefore, if the maximum value of transmission efficiency is set to $T_{4(N)}=1$, the $N$-order equation of "$f$-$f_0$" is obtained by simplification. There are at the most $N$ real solutions to this equation, i.e. there are at the most $N$ formants in the transmission spectrums of $T_{4(N)}$, whose positions are determined by $\tau_0$, $\tau_1$, $\tau_2$, $\tau_a$ and $\mu$.

Next, the laws and reasons of producing multiple formants are explored. According to the $T_{4(N)}$ expression, when $N \geq 2$, the influence term of interaction factor $\mu$ between the nearest-neighbor cavities appears in Eq. (5). Therefore, it is assumed that the coupling distance is far and there are enough transmission media between the nearest-neighbor cavities. In this case, the coupling form between the nearest-neighbor cavities is evanescent wave, so that $\mu \to 0$, then the Eq. (5) changes to

$$T_{4(N)} = \left|\frac{2^N/(\tau_1^{1/2}\tau_2^{1/2}\tau_a^{N-1})\left[j2\pi(f-f_0)+1/\tau_0+1/\tau_1+1/\tau_a\right]^{-1}}{\left[j2\pi(f-f_0)+1/\tau_0+1/\tau_2+1/\tau_a\right]\left[j2\pi(f-f_0)+1/\tau_0+2/\tau_a\right]^{N-2}}\right|, \quad (12)$$

where $N \geq 2$. According to Eq. (12), the numerator order is the same as the denominator order in $T_{4(N)}$ expression. Hence, the transmission spectrums with $T_{4(N)}$ expression only possess a single formant, which is located at the frequency point $f_0$, and the peak value of transmission efficiency is related to $\tau_0$, $\tau_1$, $\tau_2$ and $\tau_a$. Furthermore, as can be seen from the denominator of Eq. (12), the larger the value of $N$ is, the sharper the formant at $f_0$ (when $f$ approaches $f_0$ from a distance, $T$ can quickly reach the maximum value) is. Therefore, as $N$ increases, the bandwidth of the formant decreases. According to the definition of $Q$ [22], this means that $Q$ will increase as $N$ increases.

In the case of ignoring the influence of $\mu$, the influence of the size relation between $\tau_1$, $\tau_2$ and $\tau_a$ on the transmission efficiency of the drop port $P_4$ is discussed. If the designed HONAF satisfies the condition $\tau_1 \neq \tau_2 \neq \tau_a$, the maximum value of $T_{4(N)}$ can be obtained from Eqs. (4) and (12), that is

$$T_{4(1)\max} = \frac{4\tau_1\tau_2}{\left(\tau_1\tau_2/\tau_0 + \tau_1 + \tau_2\right)^2}, \quad (13)$$

$$T_{4(N)\max} = \frac{4^N \tau_1 \tau_2 \tau_a^2 \left(\tau_a/\tau_0 + 2\right)^{4-2N}}{\left(\tau_1\tau_a/\tau_0 + \tau_1 + \tau_a\right)^2 \left(\tau_2\tau_a/\tau_0 + \tau_2 + \tau_a\right)^2} \quad (N \geq 2). \quad (14)$$

In the general structure, the values of $\tau_1$, $\tau_2$ and $\tau_a$ are diminutive, i.e. the relation $1 > T_{4(1)\max} > T_{4(N)\max}$ ($N \geq 2$) is valid, indicating that the transmission efficiency of the drop port $P_4$ will decrease with the increase of $N$ without ignoring the influence of $\tau_0$. If the influence of $\tau_0$ is ignored in cascaded cavities, it can be obtained

$$T_{4(1)\max} = \frac{4\tau_1\tau_2}{\left(\tau_1 + \tau_2\right)^2}, \quad (15)$$

$$T_{4(N)\max} = \frac{16\tau_1\tau_2\tau_a^2}{\left(\tau_1 + \tau_a\right)^2 \left(\tau_2 + \tau_a\right)^2} \quad (N \geq 2). \quad (16)$$

Obviously, there is $1 > T_{4(1)\max} > T_{4(2)\max} = T_{4(N)\max}$ ($N \geq 2$), i.e. in the case of $N \geq 2$, $T_{4(N)\max}$ is a constant.

If the designed HONAF satisfies the condition (which can be satisfied if the leaky components [17] of cavities are completely symmetrical),

$$\tau_1 = \tau_2 = \tau_a, \quad (17)$$

and if the influence of $\tau_0$ is ignored, then, it can be seen from Eqs. (15) and (16) that $T_{4(N)\max} \equiv 1$ is valid obviously. Specific structure parameters are substituted into Eqs. (4) and (12), then the transmission spectra of the HONAF can be obtained (see Section II of the Supplemental Material [56]), that intuitively shows the correctness of the theory mentioned above. Therefore, a filter with high $Q$ can be obtained by cascaded cavities without changing the resonance frequency of the single isolated cavity. Moreover, the above studies found that only when $\tau_1 = \tau_2 = \tau_a$, the maximum drop efficiency of the HONAF could reach 100%.

Next, the quality factor expression ($Q_N$) of the $N^{\text{th}}$-order HONAF, in the case of $\tau_1 = \tau_2 = \tau_a$, $\tau_0 = 0$ and $\mu = 0$, is derived (see Section III of the Supplemental Material [56])

$$Q_N = Q_0 \left(2^{1/N} - 1\right)^{-1/2} \quad (N \geq 1), \quad (18)$$

where $Q_0$ is the quality factor of the single isolated cavity. Obviously, $Q_N$ is a function of $N$, when $N \to \infty$, $Q_N \to \infty$. However, in the actual situation (with energy loss), as $N$ increases, the loss of the whole CCC system will inevitably increase, hence $Q_N$ cannot approach infinity. The first derivative of $Q_N$ with respect to $N$ is $Q'_N > 0$, i.e. $Q_N$ increases monotonically with the increase of $N$. The second derivative $Q''_N < 0$, i.e. the line spectrum of $Q_N$ is upward convex type. Therefore, as $N$ increases, the growth rate of $Q_N$ becomes smaller and smaller. The cascade of more cavities will no longer play a leading role. However, in this case, $N$ is a large value, and the self-loss of the CCC system will play a leading role. The $Q_0$ of a single cavity is substituted into Eq. (18) to obtain the spectrum of $Q_N$ changing with $N$ in the CCC system, which is consistent with the above analysis (see Section III of the Supplemental Material [56]).

If we consider the loss of the cavities themselves, the larger of the loss, the smaller the number of cavities that can cascade. Another factor that determines the number of cascaded cavities is the interaction $\mu$ between cavities. Considering that it is impossible to achieve $\mu \to 0$ in the actual cascaded cavities ($\mu \neq 0$ will lead to frequency splitting), combined with our simulation data, the up-bound of $Q$ in our design is at the level of $1 \times 10^6$ (see Section III of the Supplemental Material [56]).

It's worth noting that the $\mu \to 0$ condition is harder to satisfy. The reason is that the interaction between cascaded cavities always exists, even if the coupling distance between the nearest-neighbor cavities is very long, there is still a weak interaction, and it can be seen from Eq. (5) that this kind of interaction increases with the increase of $N$ (In the denominator of the $T_{4(N)}$ expression, the influence of $\mu$ is strengthened with the increase of $N$). Hence, for the higher-order HONAF, the influence of interaction between cascaded cavities cannot be overcome theoretically. If the coupling distance between the cavities is too large, new problems will arise, such as the coupling efficiency between the waveguide channels and the cavities will decrease, thus reducing the drop efficiencies.

As the coupling distance between the cascaded cavities is reduced, the coupling form between the nearest-neighbor cavities will gradually change from evanescent wave coupling to near-field coupling. However, the discussion in this paper is limited to the category of evanescent wave coupling, and it is assumed that the HONAF still satisfies the weak coupling condition. The interaction factor $\mu$ will play a leading role in this process, so that the CCC system will show more excellent characteristics. According to Eq. (11), the change of $\mu$ leads to the change of $\tau_a$ (the specific change depends on the specific structure). From Eq. (5), it can be seen that the $\mu$ will have an impact on the transmission efficiency of the drop port, which will be intensified with the increase of $N$ (In the denominator of Eq. (5), the influence of $\mu$ polynomial will increase). Once $\mu \neq 0$, $\mu$ polynomial will cause transmission spectrums of the $T_{4(N)}$ to split from a single formant into multiple formants. When $\mu$ is slightly greater than zero, peak spikes appear in the single formant in the $T_{4(N)}$ transmission spectrums. When $\mu$ is large enough, the single formant will completely split into multiple formants in the $T_{4(N)}$ transmission spectrums. As long as $\mu \neq 0$, the splitting effect will persist. It can also be seen from Eq. (5) that the number of formants generated by the influence $\mu$ is $N$, and the positions of the formants after splitting are symmetrical about $f_0$. According to Eq. (5), the obtain the transmission spectrums of the HONAF with specific structural parameters shown that the conclusion is consistent with the above. Under the same parameters, when $N$ is large, single formant is more likely to split into multiple formants (see Section IV of the Supplemental Material [56]). This means that in the actual HONAF, even if the coupling distance between the nearest-neighbor cavities is large, as $N$ increases, its transmission spectrums will inevitably split from single formant to multiple formants.

What will be the effect of the fluctuation of the coupling strength among each cavity? After analysis (see Section IV of the Supplemental Material [56]), we found that the different coupling strength between the cavities not only affects the situation of the frequency splitting, shifts each resonance frequency point, but also reduces the transmission efficiency of the filter. This will affect the excellent performance and stability of the filter. Therefore, in practice, the coupling strength between the cavities should be as consistent as possible. It is particularly important to note that the size and material of the scatterers must be consistent, and they must be manufactured in strict accordance with the geometric arrangement rules of the PCs. In addition, when fabricating the absorbing boundary, it is necessary to isolate external disturbances.

It is important to note that when the coupling distance between the cascaded cavities is too close, the strong coupling will make the interaction between the cascaded cavities becoming more complex, even the cascaded cavities have merged into a whole, the property of the isolated cavity will gradually disappear, to create a new cavity, the emergence of new properties, such as different resonance frequency from that of isolated cavity, the number of formants is no longer equal to $N$, and the resonance frequency is not symmetrical about $f_0$. The above theory is no longer universally applicable in this case.

Furthermore, the transmission efficiency equations (Eqs. (1) to (5)) of the HONAF discussed above is also generally applicable to CROWs. Simply remove the two waveguide channels (the bus waveguide and the add/drop waveguide) or change the transmission direction of the two waveguides to coincide with that of the cascaded cavities (i.e. to be couplers for input and output ports). Equation (5) can be used to research the transmission characteristics of the CROWs (see Section V of the Supplemental Material [56]).

### III. A HONAF VERIFYING THE t-CMT

In this Section, a HONAF is designed to verify the correctness of the theory above, which can also be applied to the actual WDM system. Two topological waveguide channels are constructed by using two-dimensional (2D) square lattice GPCs and decagonal Penrose-type PQCs, and the decagonal Penrose-type PQCs is used as the cavity to obtain the four-port system of waveguide - CCC - waveguide.

The decagonal Penrose-type PQCs is shown in Fig. 2(a) [55]. The yellow scatterers in Fig. 2(a) are selected as the basic cascaded unit (isolated cavity) to construct the HONAF with cascaded cavities, as shown in Fig. 2(b) (In order to prove the correctness of the theoretical prediction in Section II and its generality with the simulation results, the basic cascaded unit S can be arbitrarily taken to construct HONAF S. The model and results can be referred to Section VI of the Supplemental Material [56]). In order to make the coupling between waveguides and cavities better meet the weak coupling condition and ensure favourable frequency selectivity, the coupling distance between waveguides and cavities is increased, i.e. more scatterers (white scatterers in Fig. 2), are taken in the coupling cavity unit adjacent to the waveguide.

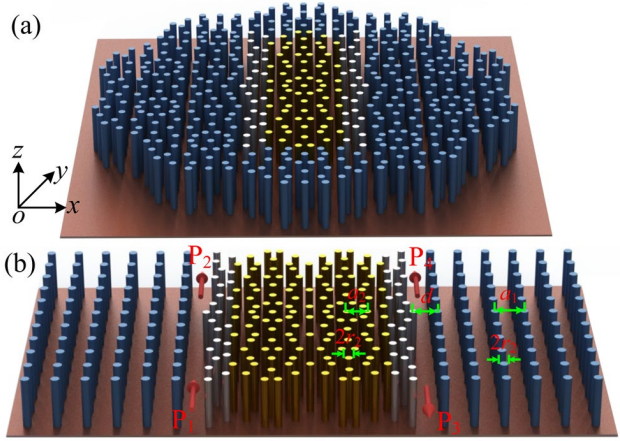

FIG. 2. (a) Decagonal Penrose-type PQCs; (b) 2nd-order HONAF model.

Ports $P_1$ and $P_2$ are provided to constitute the bus waveguide, while ports $P_3$ and $P_4$ constitute the add/drop waveguide. If the signal source is placed on port $P_1$, and external dc magnetic field in the direction of $-z$ and $+z$ is applied to the left and right GPCs respectively (the nonreciprocal directions are counterclockwise and clockwise respectively), as shown in Fig. 2(b), so that the port $P_4$ is the drop channel. If $P_3$ is taken as the input port, and the direction of the external dc magnetic field in the GPCs is not changed, the HONAF will realize the function of adding the signal and input the signal into the bus.

In the GPCs (the blue medium columns in Fig. 2(b)), the scatterer is yttrium-iron-garnet (YIG) with permittivity $\varepsilon_1=15\varepsilon_0$, and permeability $\mu_0$ in the absence of external dc magnetic field. The initial structure parameters are set as that, lattice constant $a_1=14$ mm, scatterer radius $r_1=0.11a_1$, and the background material is air. In the PQCs (the yellow and white medium columns in Fig. 2(b)), the scatterer is alumina with permittivity $\varepsilon_2=10\varepsilon_0$, and permeability $\mu_0$. The lattice constant and the scatterer radius are $a_2=10$ mm and $r_2=0.15a_2$ respectively, and the background material is also air. The coupling channel width between the GPCs and the PQCs is $d=0.875a_1$. When the external dc magnetic field is 1600 G and the frequency is 4.28 GHz [34,35,41], the permeability of the GPCs is the matrix tensor in the form of

$$\mu = \begin{bmatrix} 14\mu_0 & \pm 12.4i\mu_0 & 0 \\ \mp 12.4i\mu_0 & 14\mu_0 & 0 \\ 0 & 0 & \mu_0 \end{bmatrix}, \quad (19)$$

Since the GPCs generate gyromagnetic anisotropy, which breaks the time-reversal symmetry [57] of the system and realizes the optical quantum Hall effect. A pair of one-way electromagnetic edge states with chiral symmetry exist on its boundary [34,58]. We calculated the bands of the GPC with square lattice (see Section VII of the Supplemental Material [56]). When a dc magnetic field is applied, a topological band gap appears between the second and third bands. Moreover, we calculate the number of states of the first-order filter to obtain the common band gap between the GPCs and the PQCs, the chiral topological edge state (waveguide channel) and the defect mode of the PQCs (as a cavity) is located in the band gap.

Next, the transmission performance of the HONAF is analyzed. With initial structure parameters (whose value obtained after structural parameters optimization) and the order $N=1$ to 6, the transmission spectrum of the HONAF can be obtained [Fig.3]. As can be seen from Fig. 3, when the HONAF is in each resonance state, $T_{2(N)}\approx 0$, $T_{4(N)}\approx 1$ and $T_{3(N)}\equiv 0$, the drop efficiency reaches almost 100%. Moreover, with the increase of $N$, more and more formants in the HONAF are generated, i.e. $\mu$ causes the resonance frequency of the isolated cavity to split, producing multiple resonance frequencies, and the number of resonance frequencies is equal to the number $N$ of cascaded cavities. In other words, the function of multiple filtering can be realized through cascaded cavities. The function of multiband transmission can be realized if it is applied to CROWs. Theoretically, the $N$ can be increased continuously to obtain the HONAF with more channels and realize the filtering function with any number of channels. It can be seen from Fig. 3 that when $N$ increases, the bandwidth of formants decreases, and the formants band width at both ends is much narrower, while the formants bandwidth in the middle is much wider. Furthermore, with the increase of $N$, the out-of-band rejection ratios of the transmission spectrums of port $P_4$ also increases. The above numerical simulation results are basically consistent with the theoretical prediction. It is worth noting that the cascaded cavities produce multiple resonance frequencies that are not symmetric with respect to $f_0$, but are randomly dispersed in the frequency interval around $f_0$, which is slightly different from the theoretical prediction. The reason is that the coupling distance between cascaded cavities is diminutive, resulting the interaction between the cascaded cavities is strong, and the interaction is not only from the nearest-neighbor cavities, but also from the next-nearest-neighbor cavities. In this case, the interaction coefficient cannot be simply described by $\mu$, more complex interaction factors need to be introduced, which is different from the original hypothesis of the theory above. Moreover, when the coupling distance between cascaded cavities is too diminutive, the t-CMT will no longer be accurate [22]. The explanation of this situation by the theory derived above is not reasonable, therefore, the numerical simulation is slightly inconsistent with the theory. In addition, by analyzing the electric field distribution laws in the cascaded cavities, it is also proved that the reason why the HONAF possesses multiple filtering function lies in the interaction between cascaded cavities from the perspective of photon localization (see Section VIII of the Supplemental Material [56]).

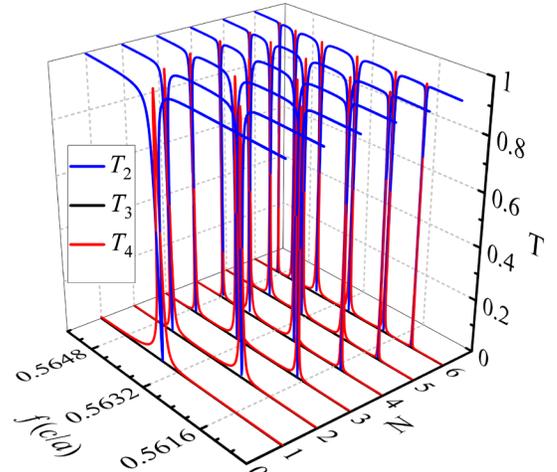

FIG. 3. Transmission spectrum of the HONAF.

Next, the quality factor distribution laws of the HONAF are analyzed. With the initial structure parameters and the order $N=1$ to 24, the $Q$s in each resonance state are obtained by numerical calculation [Fig. 4]. As can be seen from Fig. 4, with the increase of $N$, the average quality factor $\bar{Q} \equiv 1/N \sum_{i=1}^{N} Q_{Ni}$ also increases, indicating that performance of the HONAF becomes better with the increase of $N$. For the HONAF with particular order, the distribution of $Q$s at each resonance state is U-shaped. The $Q$s are the largest one at the resonance state of both left and right ends, while it is the smallest one at the intermediate resonance state (e.g. for the 3$^{rd}$-order HONAF, $f_{032}$ is called the intermediate resonance state, while $f_{031}$ and $f_{033}$ are respectively called the left and right resonance states, and the other order HONAFs are similar). Moreover, the distribution laws of $Q$s at each resonant state are also consistent with the photon localization laws in the cascaded cavities mentioned above (see Section VIII of the Supplemental Material [56]). At the intermediate resonance state, the photons are concentrated in the cavities at the left and right ends, which are easy to leak into the waveguide channels, hence the $Q$s are much smaller. When the cascaded cavities are at the resonance state at the left and right ends, the photons are concentrated in the intermediate cavities, which are not easy to leak, the $Q$s are naturally much larger. This is also the reason why the $Q$s increases rapidly with the increase of $N$ (it is more and more difficult for photons to leak out) at the resonance state of both left and right ends of the HONAF, while the $Q$s increases slowly at the intermediate resonance state.

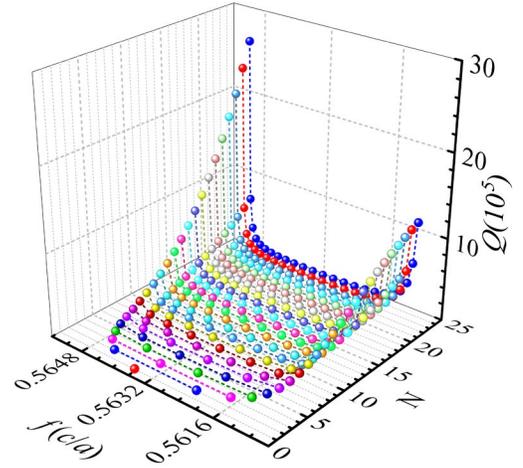

FIG. 4. The $Q$ of HONAF in each resonance state.

It is worth noting that the drop channels of the HONAF can be regulated by external dc magnetic field, i.e. the direction of the external dc magnetic field can be controlled to adjust the nonreciprocal direction of the add/drop waveguide, which determines the drop channels. For the HONAF designed above, change the external dc magnetic field direction of GPCs in the right half to -$z$, then the drop channel becomes $P_3$. With the original structural parameters, the transmission spectrums and the distribution diagrams of electric field modulus value at resonance state in 5$^{th}$-order HONAF with defects or not are shown in Fig. 5. When the external dc magnetic field direction of the right half part of GPCs is -$z$, $T_{3(1)}\approx1$ and $T_{2(1)}\approx0$, the drop efficiency is also almost 100%. Meanwhile, $T_{4(1)}\equiv0$ due to the nonreciprocal waveguide. By comparing Figs. 5(a) and 5(b), it can be seen that $P_3$ or $P_4$ is used as the drop channel, transmission efficiencies of the HONAF are basically the same. Therefore, changing the drop port will not affect the performance of the HONAF. Moreover, the nonreciprocal waveguide channels designed possess a topological nontrivial state, i.e. when there are defects in the HONAF waveguides, the light wave in the waveguides can bypass the defects and achieve 100% forward transmission. It can be found from Figs. 5(c), 5(d) and 5(e) that, light wave can bypass the defects and transmit forward without loss, indicating that waveguide channels in the HONAF is topologically protected.

All the numerical simulation data are obtained by the finite element method. In all the filter model simulations, only electromagnetic waves with TM mode are considered. The boundary in the $z$ direction of the model is defaulted to be infinite, and the boundary in the $x$ and $y$ directions uses a perfectly matched layer as the absorbing boundary, on which the electromagnetic wave can quickly decay to zero. The 2D PCs generally use 3D finite height structures for experimental design in practice. The filter designed works at the microwave frequency band, and the microwave experiment of GPCs has been reported [19,35]. Therefore, a simple experimental implementation scheme can be designed to clarify the feasibility of the experiment and deal with the problems that may be encountered during the experiment (see Section IX of the Supplemental Material [56]).

## IV. CONCLUSION

In this paper, the t-CMT is used to derive the transmission efficiency equations of the CCC system, which is applicable to the general situation. A design scheme of universal HONAF is proposed, which can improve the $Q$ and $T$ through cascaded cavities and realize the function of multi-channel filtering. It is concluded that the interaction between cascaded cavities leads to the splitting of transmission spectrums from a single formant to multiple formants. In the case of $\mu=0$, $\tau_0=0$ and $\tau_1=\tau_2=\tau_a$, the change equation of the $Q_N$ of the single formant HONAF increases with the increase of $N$ is derived. A HONAF which is composed of GPCs and the decagonal Penrose-type PQCs is designed to work at microwave frequency band. With the change of $N$, the transmission characteristics and the change laws of $Q$s of the HONAF are obtained, which verifies the consistency between theoretical prediction and numerical simulation. By adjusting the direction of the external dc magnetic field, the convertibility of drop channel of the HONAF is verified, and the topological protec-tion of the waveguide channels is also verified by introducing point defects (size and material of two scatterers changed) and line defects.

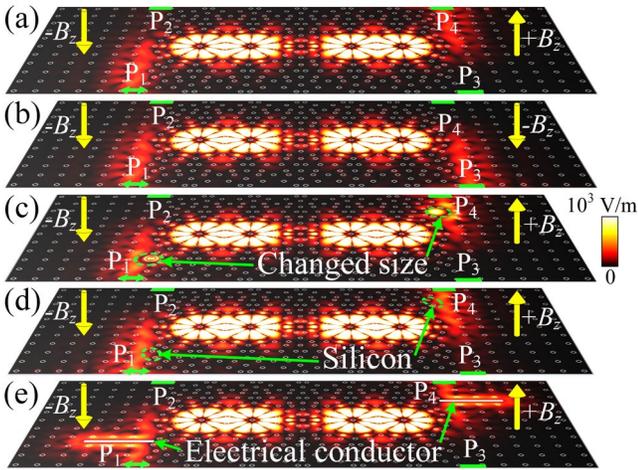

FIG. 5. The distribution diagrams of electric field modulus value at resonance state $f_{052}$ in 5$^{th}$-order HONAF in the different cases: (a) and (b), the external dc magnetic field direction of GPCs in the right half part is +$z$ and -$z$, respectively; (c), (d) and (e), reduce and increase the radius of the scatterer as indicated by the green dotted circle, replace the quasicrystal material alumina with silicon marked by green ellipse dotted box and an ideal electrical conductor marked by the white line, respectively. The green double arrow represents the light source and the green thick line represents the monitors.

ACKNOWLEDGMENT

This work was supported by the National Natural Science Foundation of China (Grant Nos. 61405058 and 62075059), the Natural Science Foundation of Hunan Province (Grant Nos. 2017JJ2048 and 2020JJ4161), and the Fundamental Research Funds for the Central Universities (Grant No. 531118040112). The authors acknowledge Prof. J. Q. Liu for software sponsorship.

REFERENCES

[1] D. Marpaung, J. Yao, and J. Capmany, Integrated microwave photonics, Nat. Photonics 13, 80 (2019).
[2] Y. Xiong, T. Umeda, X. Zhang, M. Morifuji, H. Kajii, A. Maruta, and M. Kondow, Photonic crystal circular-defect microcavity laser designed for wavelength division multiplexing, IEEE J. Sel. Top. Quantum Electron. 24, 1 (2018).
[3] X. Wang, Z. Pang, H. Yang, and Y. Qi, Theoretical study of subwavelength circular grating fabrication based on continuously exposed eurface plasmon interference lithography, Results Phys. 14, 102446 (2019).
[4] H. Tong, Y. Xu, Y. Su, and X. Wang, Theoretical study for fabricating elliptical subwavelength nanohole arrays by higher-order waveguide-mode interference, Results Phys. 14, 102460 (2019).
[5] E. Liu, W. Tan, B. Yan, J. Xie, R. Ge, and J. Liu, Broadband ultra-flattened dispersion, ultra-low confinement loss and large effective mode area in an octagonal photonic quasi-crystal fiber, J. Opt. Soc. Am. A 35, 431 (2018).
[6] Q. Liu, B. Yan, and J. Liu, U-shaped photonic quasi-crystal fiber sensor with high sensitivity based on surface plasmon resonance, Appl. Phys. Express 12, 052014 (2019).
[7] E. Liu, S. Liang, and J. Liu, Double-cladding structure dependence of guiding characteristics in six-fold symmetric photonic quasi-crystal fiber, Superlattices Microstruct. 130, 61 (2019).
[8] Z. Huo, E. Liu, and J. Liu, Hollow-core photonic quasi-crystal fiber with high birefringence and ultra-low nonlinearity, Chinese Opt. Lett. 18, 030603 (2020).
[9] J. Liu and Z. Fan, Size limits for focusing of two-dimensional photonic quasicrystal lenses, IEEE Photonics Technol. Lett. 30, 1001 (2018).
[10] H. Zhao, J. Xie, and J. Liu, An approximate theoretical explanation for super-resolution imaging of two-dimensional photonic quasi-crystal flat lens, Appl. Phys. Express 13, 022007 (2020).
[11] M. Menotti, B. Morrison, K. Tan, Z. Vernon, J. E. Sipe, and M. Liscidini, Nonlinear coupling of linearly uncoupled resonators, Phys. Rev. Lett. 122, 013904 (2019).
[12] P. Yu, X. Chen, Z. Yi, Y. Tang, H. Yang, Z. Zhou, T. Duan, S. Cheng, J. Zhang, and Y. Yi, A numerical research of wideband solar absorber based on refractory metal from visible to near infrared, Opt. Mater. (Amst). 97, 109400 (2019).
[13] C. Liang, Y. Zhang, Z. Yi, X. Chen, Z. Zhou, H. Yang, Y. Yi, Y. Tang, W. Yao, and Y. Yi, A broadband and polarization-independent metamaterial perfect absorber with monolayer Cr and Ti elliptical disks array, Results Phys. 15, 102635 (2019).
[14] M. Li, C. Liang, Y. Zhang, Z. Yi, X. Chen, Z. Zhou, H. Yang, Y. Tang, and Y. Yi, Terahertz wideband perfect absorber based on open loop with cross nested structure, Results Phys. 15, 102603 (2019).
[15] C. Liang, Z. Yi, X. Chen, Y. Tang, Y. Yi, Z. Zhou, X. Wu, Z. Huang, Y. Yi, and G. Zhang, Dual-band infrared perfect absorber based on a Ag-dielectric-Ag multi-layer films with nanoring grooves arrays, Plasmonics 15, 93 (2020).
[16] F. Morichetti, C. Ferrari, A. Canciamilla, and A. Melloni, The first decade of coupled resonator optical waveguides: bringing slow light to applications, Laser Photon. Rev. 6, 74 (2012).
[17] K. Maeno, Y. Takahashi, T. Nakamura, T. Asano, and S. Noda, Analysis of high-Q photonic crystal L3 nanocavities designed by visualization of the leaky components, Opt. Express 25, 367 (2017).
[18] S. Combrié, G. Lehoucq, G. Moille, A. Martin, and A. De Rossi, Comb of high-Q resonances in a compact photonic cavity, Laser Photon. Rev. 11, 1700099 (2017).
[19] F.-F. Li, H.-X. Wang, Z. Xiong, Q. Lou, P. Chen, R.-X. Wu, Y. Poo, J.-H. Jiang, and S. John, Topological light-trapping on a dislocation, Nat. Commun. 9, 2462 (2018).
[20] M. Clementi, K. Debnath, M. Sotto, A. Barone, A. Z. Khokhar, T. D. Bucio, S. Saito, F. Y. Gardes, D. Bajoni, and M. Galli, Cavity-enhanced harmonic generation in silicon rich nitride photonic crystal microresonators, Appl. Phys. Lett. 114, 131103 (2019).
[21] Y. Ota, F. Liu, R. Katsumi, K. Watanabe, K. Wakabayashi, Y. Arakawa, and S. Iwamoto, Photonic crystal nanocavity based on a topological corner state, Optica 6, 786 (2019).
[22] J. D. Joannopoulos, S. G. Johnson, J. N. Winn, and R. D. Meade, *Photonic crystals: molding the flow of light*, Princeton University Press, New Jersey, USA, 2008.
[23] F. Bessin, A. M. Perego, K. Staliunas, S. K. Turitsyn, A. Kudlinski, M. Conforti, and A. Mussot, Gain-through-filtering enables tuneable frequency comb generation in passive optical resonators, Nat. Commun. 10, 4489 (2019).
[24] G. Frigenti, L. Cavigli, A. Fernández-Bienes, F. Ratto, S. Centi, T. García-Fernández, G. Nunzi Conti, and S. Soria, Resonant microbubble as a microfluidic stage for

all-optical photoacoustic sensing, Phys. Rev. Appl. 12, 014062 (2019).

[25] X. Liu, D. Popa, and N. Akhmediev, Revealing the transition dynamics from Q switching to mode locking in a soliton laser, Phys. Rev. Lett. 123, 093901 (2019).

[26] K. Tao, J.-J. Xiao, and X. Yin, Nonreciprocal photonic crystal add-drop filter, Appl. Phys. Lett. 105, 211105 (2014).

[27] J. Zhou, W. Han, Y. Meng, H. Song, D. Mu, and W. Park, Optical properties of direct and indirect coupling of cascaded cavities in resonator-waveguide systems, J. Light. Technol. 32, 4104 (2014).

[28] N. Li, Z. Su, Purnawirman, E. Salih Magden, C. V. Poulton, A. Ruocco, N. Singh, M. J. Byrd, J. D. B. Bradley, G. Leake, and M. R. Watts, Athermal synchronization of laser source with WDM filter in a silicon photonics platform, Appl. Phys. Lett. 110, 211105 (2017).

[29] G. Cantarella, C. Klitis, M. Sorel, and M. J. Strain, Silicon photonic filters with high rejection of both TE and TM modes for on-chip four wave mixing applications, Opt. Express 25, 19711 (2017).

[30] M. Mendez-Astudillo, H. Okayama, and H. Nakajima, Silicon optical filter with transmission peaks in wide stopband obtained by anti-symmetric photonic crystal with defect in multimode waveguides, Opt. Express 26, 1841 (2018).

[31] E. S. Magden, N. Li, M. Raval, C. V. Poulton, A. Ruocco, N. Singh, D. Vermeulen, E. P. Ippen, L. A. Kolodziejski, and M. R. Watts, Transmissive silicon photonic dichroic filters with spectrally selective waveguides, Nat. Commun. 9, 3009 (2018).

[32] M. Tokushima, Linear spectral response of a Fano-resonant graded-stub filter based on pillar-photonic-crystal waveguides, Opt. Lett. 43, 431 (2018).

[33] C. W. Peterson, W. A. Benalcazar, T. L. Hughes, and G. Bahl, A quantized microwave quadrupole insulator with topologically protected corner states, Nature 555, 346 (2018).

[34] Z. Wang, Y. D. Chong, J. D. Joannopoulos, and M. Soljačić, Reflection-free one-way edge modes in a gyromagnetic photonic crystal, Phys. Rev. Lett. 100, 013905 (2008).

[35] Z. Wang, Y. Chong, J. D. Joannopoulos, and M. Soljačić, Observation of unidirectional backscattering-immune topological electromagnetic states, Nature 461, 772 (2009).

[36] L.-H. Wu and X. Hu, Scheme for achieving a topological photonic crystal by using dielectric material, Phys. Rev. Lett. 114, 223901 (2015).

[37] J.-W. Dong, X.-D. Chen, H. Zhu, Y. Wang, and X. Zhang, Valley photonic crystals for control of spin and topology, Nat. Mater. 16, 298 (2017).

[38] Y. Fang and Y. Zhang, Perfect nonreciprocal absorption based on metamaterial slab, Plasmonics 13, 661 (2018).

[39] X.-Y. Zhu, S. K. Gupta, X.-C. Sun, C. He, G.-X. Li, J.-H. Jiang, X.-P. Liu, M.-H. Lu, and Y.-F. Chen, Z2 topological edge state in honeycomb lattice of coupled resonant optical waveguides with a flat band, Opt. Express 26, 24307 (2018).

[40] X.-D. Chen, W.-M. Deng, F.-L. Zhao, and J.-W. Dong, Accidental double dirac cones and robust edge states in topological anisotropic photonic crystals, Laser Photon. Rev. 12, 1800073 (2018).

[41] J. Chen, W. Liang, and Z.-Y. Li, Strong coupling of topological edge states enabling group-dispersionless slow light in magneto-optical photonic crystals, Phys. Rev. B 99, 014103 (2019).

[42] W. Song, W. Sun, C. Chen, Q. Song, S. Xiao, S. Zhu, and T. Li, Robust and broadband optical coupling by topological waveguide arrays, Laser Photon. Rev. 14, 1900193 (2020).

[43] B. Yan, J. Xie, E. Liu, Y. Peng, R. Ge, J. Liu, and S. Wen, Topological edge state in the two-dimensional Stampfli-triangle photonic crystals, Phys. Rev. Appl. 12, 044004 (2019).

[44] T. Hou, R. Ge, W. Tan, and J. Liu, One-way rotating state of multi-periodicity frequency bands in circular photonic crystal, J. Phys. D. Appl. Phys. 53, 075104 (2020).

[45] J. W. You, Z. Lan, Q. Bao, and N. C. Panoiu, Valley-hall topological plasmons in a graphene nanohole plasmonic crystal waveguide, IEEE J. Sel. Top. Quantum Electron. 26, 1 (2020).

[46] Y. Peng, B. Yan, J. Xie, E. Liu, H. Li, R. Ge, F. Gao, and J. Liu, Variation of topological edge states of 2D honeycomb lattice photonic crystals, Phys. Status Solidi – Rapid Res. Lett. 14, 2000202 (2020).

[47] B.-Y. Xie, H.-F. Wang, H.-X. Wang, X.-Y. Zhu, J.-H. Jiang, M.-H. Lu, and Y.-F. Chen, Second-order photonic topological insulator with corner states, Phys. Rev. B 98, 205147 (2018).

[48] X.-D. Chen, W.-M. Deng, F.-L. Shi, F.-L. Zhao, M. Chen, and J.-W. Dong, Direct observation of corner states in second-order topological photonic crystal slabs, Phys. Rev. Lett. 122, 233902 (2019).

[49] Y. Liu, S. Leung, F.-F. Li, Z.-K. Lin, X. Tao, Y. Poo, and J.-H. Jiang, Bulk-disclination correspondence in topological crystalline insulators, Nature 589, 381 (2021).

[50] M. A. Bandres, S. Wittek, G. Harari, M. Parto, J. Ren, M. Segev, D. N. Christodoulides, and M. Khajavikhan, Topological insulator laser: experiments, Science 359, eaar4005 (2018).

[51] R. Ge, B. Yan, J. Xie, E. Liu, W. Tan, and J. Liu, Logic gates based on edge states in gyromagnetic photonic crystal, J. Magn. Magn. Mater. 500, 166367 (2020).

[52] Y. Wang, D. Zhang, B. Xu, Z. Dong, X. Zeng, J. Pei, S. Xu, and Q. Xue, Four ports double Y-shaped ultrawideband magneto-photonic crystals circulator for 5G communication system, IEEE Access 7, 120463 (2019).

[53] S. Ma and S. M. Anlage, Microwave applications of photonic topological insulators, Appl. Phys. Lett. **116**, 250502 (2020).
[54] Y. S. Chan, C. T. Chan, and Z. Y. Liu, Photonic band gaps in two dimensional photonic quasicrystals, Phys. Rev. Lett. **80**, 956 (1998).
[55] Z. Fan, J. Liu, S. Chen, H. Chang, C. Guan, and L. Yuan, Comparative study of photonic band gaps of germanium-based two-dimensional triangular-lattice and square-lattice and decagonal quasi-periodic photonic crystals, Microelectron. Eng. **96**, 11 (2012).
[56] See Supplemental Material at http://link.aps.org/supplemental/10.1103/PhysRevB.××.×××××× for supplemental theoretical derivation, physical model, numerical calculations, experimental design, and figures.
[57] F. D. M. Haldane and S. Raghu, Possible realization of directional optical waveguides in photonic crystals with broken time-reversal symmetry, Phys. Rev. Lett. **100**, 013904 (2008).
[58] Y. Hatsugai, Chern number and edge states in the integer quantum Hall effect, Phys. Rev. Lett. **71**, 3697 (1993).

# Supplemental Material

# High-Order Nonreciprocal Add-Drop Filter

**Hang Li, Rui Ge, Yuchen Peng, Bei Yan, Jianlan Xie, Jianjun Liu\*, and Shuangchun Wen**

*Key Laboratory for Micro/Nano Optoelectronic Devices of Ministry of Education & Hunan Provincial Key Laboratory of Low-Dimensional Structural Physics and Devices, School of Physics and Electronics, Hunan University, Changsha 410082, China*

### I. Transmission efficiency equations of the HONAF

The following equations are derived to predict the transmission efficiency of HONAF with cascaded cavities. According to the t-CMT [SF1], [SF2], in the cavities and waveguides, the evolution equations of the normalized amplitude of the fields over time are

$$dA_{N1}/dt = \left(j2\pi f_0 - 1/\tau_0 - 1/\tau_1 - 1/\tau_a\right) A_{N1} - j\mu A_{N2} + \sqrt{2/\tau_1} S_1, \quad (S1)$$

$$dA_{Ni}/dt = \left(j2\pi f_0 - 1/\tau_0 - 2/\tau_a\right) A_{Ni} - j\mu A_{N(i-1)} - j\mu A_{N(i+1)} + \sqrt{2/\tau_a} S_{N(i-1)} \quad (i=2,3,\cdots,N-1), \quad (S2)$$

$$dA_{NN}/dt = \left(j2\pi f_0 - 1/\tau_0 - 1/\tau_2 - 1/\tau_a\right) A_{NN} - j\mu A_{N(N-1)} + \sqrt{2/\tau_a} S_{N(N-1)}, \quad (S3)$$

$$S_{Ni} = \sqrt{2/\tau_a} A_{Ni} \quad (i=1,2,\cdots,N-1), \quad (S4)$$

$$S_{NN} = \sqrt{2/\tau_2} A_{NN}. \quad (S5)$$

Suppose port P$_4$ is the drop channel, according to the nonreciprocity of waveguides, $S_3$=0, then,

$$S_4 = S_{NN}. \quad (S6)$$

According to the law of conservation of energy,

$$S_2 = S_1 - S_4. \quad (S7)$$

The physical significances of all the parameters are consistent with the text. In addition, the interaction between the next-nearest neighbor cavities has been ignored by the (S1) to (S3). When the coupling distance between the next-nearest neighbor cavities is far, and after the isolation of the intermediate cavities, this kind of neglect is allowed.

The input time harmonic fields, the HONAF system is a linear system, whose frequency is conserved. Therefore, if the mixed frequency of the fields input at port $P_1$ is $f$, the fields oscillate in the system in the form of $e^{j2\pi ft}$, and the result can be obtained

$$dA_{Ni}/dt = j2\pi f A_{Ni} \quad (i=1,2,\cdots N). \tag{S8}$$

Substituting (S4), (S5) and (S8) into the (S1) to (S3), and simplify the terms to get

$$\left[j2\pi(f-f_0)+1/\tau_0+1/\tau_1+1/\tau_a\right]A_{N1} = -j\mu A_{N2} + \sqrt{2/\tau_1}S_1, \tag{S9}$$

$$\left[j2\pi(f-f_0)+1/\tau_0+2/\tau_a\right]A_{Ni} = \left(-j\mu+2/\tau_a\right)A_{N(i-1)} - j\mu A_{N(i+1)} \quad (i=2,3,\cdots,N-1), \tag{S10}$$

$$\left[j2\pi(f-f_0)+1/\tau_0+1/\tau_2+1/\tau_a\right]A_{NN} = \left(-j\mu+2/\tau_a\right)A_{N(N-1)}. \tag{S11}$$

Introduce parameters $a$, $b$, $c$, $d$, and set as

$$a = j2\pi(f-f_0)+1/\tau_0+1/\tau_1+1/\tau_a, \tag{S12}$$

$$b = j2\pi(f-f_0)+1/\tau_0+2/\tau_a, \tag{S13}$$

$$c = j\mu(-j\mu+2/\tau_a), \tag{S14}$$

$$d = j2\pi(f-f_0)+1/\tau_0+1/\tau_2+1/\tau_a. \tag{S15}$$

Substituting (S12) to (S15) into (S9) to (S11), the expression of $A_{N1}$ can be solved from (S9). In (S10), set $i=2$ and substitute the expression of $A_{N1}$ into (S10) to solve the expression of $A_{N2}$. Then set $i=3$ in (S10), and substitute the expression of $A_{N2}$ into the (S10) to solve the expression of $A_{N3}$. Repeat this process to get intermediate equations,

$$aA_{N1} = -j\mu A_{N2} + \sqrt{2/\tau_1}S_1, \tag{S16}$$

$$(ab+c)A_{N2} = -j\mu a A_{N3} + \sqrt{2/\tau_1}\left(-j\mu+2/\tau_a\right)S_1, \tag{S17}$$

$$\left[(ab+c)b+ac\right]A_{N3} = -j\mu(ab+c)A_{N4} + \sqrt{2/\tau_1} \times \left(-j\mu+2/\tau_a\right)^2 S_1, \tag{S18}$$

$$\left\{\left[(ab+c)b+ac\right]b+(ab+c)c\right\}A_{N4} = -j\mu\left[(ab+c)b+ac\right] \times A_{N5} + \sqrt{2/\tau_1}\left(-j\mu+2/\tau_a\right)^3 S_1, \tag{S19}$$

…

The coefficients of $A_{Ni}$ in (S16) to (S19) possess an iterative relation, which is actually determined by the relation between (S1) to (S5), and the iterative relation is always valid when $i \leq N-1$. Therefore, the intermediate equations set as

$$H_n A_{Nn} = -j\mu F_n A_{N(n+1)} + \sqrt{2/\tau_1} \left(-j\mu + 2/\tau_a\right)^{n-1} S_1$$
$$(n = 1, 2, \cdots, N-1)$$
. (S20)

$H_n$ and $F_n$ are defined as sequence of functions of $n$, where $n$ is positive integer, and $n_{max}=N-1$. From the (S16) to (S19), can get

$$H_1 = a,$$ (S21)

$$H_2 = bH_1 + c,$$ (S22)

$$H_3 = bH_2 + cH_1,$$ (S23)

$$H_4 = bH_3 + cH_2,$$ (S24)

$$F_1 = 1,$$ (S25)

$$F_2 = a,$$ (S26)

$$F_3 = bF_2 + c,$$ (S27)

$$F_4 = bF_3 + cF_2.$$ (S28)

The general term formulas of $H_n$ and $F_n$ can be obtained from (S21) to (S28). The correctness of this recursive method is proved by the original (S1) to (S5). Therefore,

$$H_{n+2} = bH_{n+1} + cH_n \quad (n = 1, 2, \cdots N-1),$$ (S29)

$$F_{n+2} = bF_{n+1} + cF_n \quad (n = 1, 2, \cdots N-1).$$ (S30)

Obviously, the recursive expressions of $H_n$ and $F_n$ are consistent. The general term formulas of $H_n$ and $F_n$ is obtained by using the method of characteristic equation. Hence, let the general term formulas of $H_n$ and $F_n$ be respectively,

$$H_n = w_1 x_1^n + w_2 x_2^n,$$ (S31)

$$F_n = v_1 x_1^n + v_2 x_2^n,$$ (S32)

where, $w_1$, $w_2$, $v_1$ and $v_2$ are the coefficients, $x_{1,2}$ is the solution of the characteristic equation, and

$$x_{1,2} = \frac{b \pm \sqrt{b^2 + 4c}}{2},$$ (S33)

$$w_1 = \frac{ab + c - ax_2}{x_1^2 - x_1 x_2}$$

$$w_2 = \frac{ab + c - ax_1}{x_2^2 - x_1 x_2}, \tag{S34}$$

$$v_1 = \frac{a - x_2}{x_1^2 - x_1 x_2}$$

$$v_2 = \frac{a - x_1}{x_2^2 - x_1 x_2}. \tag{S35}$$

By substituting (S34) and (S35) into (S31) and (S32) respectively, the general term formulas of $H_n$ and $F_n$ are obtained as

$$H_n = \frac{(ab + c - ax_2) x_1^{n-1}}{x_1 - x_2} + \frac{(ab + c - ax_1) x_2^{n-1}}{x_2 - x_1}, \tag{S36}$$

$$F_n = \frac{(a - x_2) x_1^{n-1}}{x_1 - x_2} + \frac{(a - x_1) x_2^{n-1}}{x_2 - x_1}. \tag{S37}$$

Substitute (S36) and (S37) back into (S20), and let $n=N-1$ ($n \geq 1 \rightarrow N \geq 2$), the expression of $A_{N(N-1)}$ can be solved. Substitute this expression into (S11), and the expression of $A_{NN}$ can be obtained as

$$A_{NN} = \frac{\sqrt{2/\tau_1} \left(-j\mu + 2/\tau_a\right)^{N-1} S_1}{dH_{N-1} + cF_{N-1}}. \tag{S38}$$

Then substitute (S38) into (S5), and utilize (S6) and (S7), then get

$$S_4 = S_{NN} = \frac{2/(\tau_1 \tau_2)^{1/2} \left(-j\mu + 2/\tau_a\right)^{N-1} S_1}{dH_{N-1} + cF_{N-1}}, \tag{S39}$$

$$S_2 = S_1 - S_4 = \left(1 - \frac{2/(\tau_1 \tau_2)^{1/2} \left(-j\mu + 2/\tau_a\right)^{N-1}}{dH_{N-1} + cF_{N-1}}\right) S_1. \tag{S40}$$

Hence, the transmission efficiencies of HONAF system is

$$T_{2(N)} = \left|\frac{S_2}{S_1}\right|^2 = \left|1 - \frac{2/(\tau_1 \tau_2)^{1/2} \left(-j\mu + 2/\tau_a\right)^{N-1}}{dH_{N-1} + cF_{N-1}}\right|^2 \quad (N \geq 2), \tag{S41}$$

$$T_{3(N)} = \left|\frac{S_3}{S_1}\right|^2 = 0 \quad (N \geq 1), \tag{S42}$$

$$T_{4(N)} = \left|\frac{S_4}{S_1}\right|^2 = \left|\frac{2/(\tau_1 \tau_2)^{1/2} \left(-j\mu + 2/\tau_a\right)^{N-1}}{dH_{N-1} + cF_{N-1}}\right|^2 \quad (N \geq 2). \tag{S43}$$

In the case of *N*=1, it's easy to get

$$T_{2(1)} = \left|1 - \frac{2/(\tau_1\tau_2)^{1/2}}{j2\pi(f-f_0)+1/\tau_0+1/\tau_1+1/\tau_2}\right|^2, \quad (S44)$$

$$T_{4(1)} = \left|\frac{2/(\tau_1\tau_2)^{1/2}}{j2\pi(f-f_0)+1/\tau_0+1/\tau_1+1/\tau_2}\right|^2. \quad (S45)$$

Now, consider the polynomial $dH_{N-1}+cF_{N-1}$. Substitute (S33) into (S36) and (S37), set $n=N-1$ ($N\geq 2$), and then substitute $dH_{N-1}+cF_{N-1}$, which is combined and simplified to get

$$dH_{N-1}+cF_{N-1} = \frac{abd+2ac+2cd-bc+(ad+c)(b^2+4c)^{1/2}}{2^{N-1}(b^2+4c)^{1/2}\left[b+(b^2+4c)^{1/2}\right]^{2-N}}$$

$$-\frac{abd+2ac+2cd-bc-(ad+c)(b^2+4c)^{1/2}}{2^{N-1}(b^2+4c)^{1/2}\left[b-(b^2+4c)^{1/2}\right]^{2-N}}. \quad (S46)$$

The expressions for $T_{4(2)}$, $T_{4(3)}$, and $T_{4(4)}$ are listed below,

$$T_{4(2)} = \Big|2/(\tau_1\tau_2)^{1/2}(-j\mu+2/\tau_a)/\{[j2\pi(f-f_0)+1/\tau_0+1/\tau_1\\+1/\tau_a][j2\pi(f-f_0)+1/\tau_0+1/\tau_2+1/\tau_a]+j\mu(-j\mu\\+2/\tau_a)\}\Big|^2, \quad (S47)$$

$$T_{4(3)} = \Big|2/(\tau_1\tau_2)^{1/2}(-j\mu+2/\tau_a)^2/\{[j2\pi(f-f_0)+1/\tau_0+1/\tau_1\\+1/\tau_a][j2\pi(f-f_0)+1/\tau_0+1/\tau_2+1/\tau_a][j2\pi(f-f_0)\\+1/\tau_0+2/\tau_a]+j\mu(-j\mu+2/\tau_a)[j4\pi(f-f_0)+2/\tau_0\\+1/\tau_1+1/\tau_2+2/\tau_a]\}\Big|^2, \quad (S48)$$

$$T_{4(4)} = \left| 2/(\tau_1\tau_2)^{1/2} (-j\mu + 2/\tau_a)^3 \Big/ \{ [j2\pi(f - f_0) + 1/\tau_0 + 1/\tau_1 \right.$$
$$+ 1/\tau_a][j2\pi(f - f_0) + 1/\tau_0 + 1/\tau_2 + 1/\tau_a][j2\pi(f - f_0)$$
$$+ 1/\tau_0 + 2/\tau_a]^2 + j\mu(-j\mu + 2/\tau_a)[j2\pi(f - f_0) + 1/\tau_0$$
$$+ 1/\tau_1 + 1/\tau_a][j2\pi(f - f_0) + 1/\tau_0 + 1/\tau_2 + 1/\tau_a] + j\mu \times$$
$$(-j\mu + 2/\tau_a)[j4\pi(f - f_0) + 2/\tau_0 + 1/\tau_1 + 1/\tau_2 + 2/\tau_a]$$
$$\left. \times [j2\pi(f - f_0) + 1/\tau_0 + 2/\tau_a] + (j\mu)^2(-j\mu + 2/\tau_a)^2 \} \right|^2$$
(S49)

### II. In the case of $\mu \to 0$, transmission spectrums of the HONAF system

When order $N=1$ to 3, according to (S45) and (S53), the transmission spectrums of all the ports of HONAF under a specific $f_0$, $\tau_1$, $\tau_2$ and $\tau_a$ (ignoring the influence of $\tau_0$) are obtained, as shown in Fig. S1.

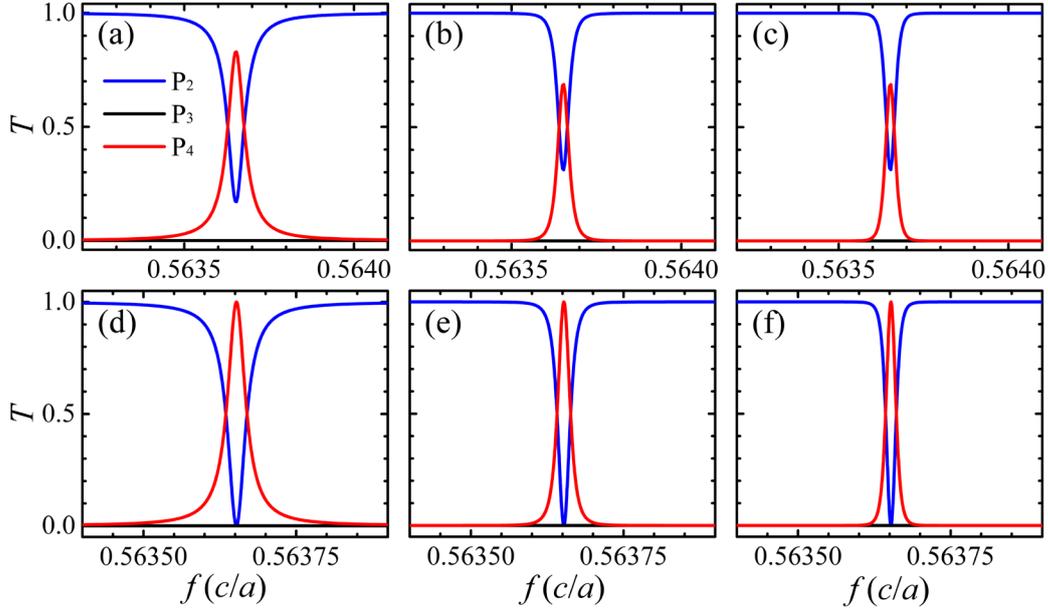

Fig. S1. Transmission spectrums of the HONAF, suppose $f_0=0.56365 \times c/a$ ($a$ is the lattice constant, $c$ is the speed of light in the vacuum, and assume $a=14$ mm, same below), $\tau_1=8.5765 \times 10^{-7}$ s, $\mu=0$. (a)-(c) respectively represent the transmission spectrums of the HONAF when $N=1,2$ and 3, and the parameter values (a) $\tau_2=3.5765 \times 10^{-7}$ s; (b)-(c) $\tau_2=\tau_a=3.5765 \times 10^{-7}$ s; (d)-(f) correspond to the transmission spectrums of the HONAF when $N=1,2$ and 3 satisfy the condition $\tau_1=\tau_2=\tau_a$.

It can be intuitively seen from Fig. S1(a) to S1(c) that when $\tau_1$, $\tau_2$ and $\tau_a$ are not the same, $1>T_{4(1)max}>T_{4(2)max}=T_{4(3)max}$. According to Fig. S1(d) to S1(f), when $\tau_1=\tau_2=\tau_a$, $T_{4(N)max}\equiv 1$. It can also be intuitively seen from Fig. S1 that the bandwidth of transmission spectrums decreases with the increase of $N$, i.e. the $Q$ of the HONAF will increase with the increase of $N$.

### III. In the case of $\tau_1=\tau_2=\tau_a$, $\tau_a=0$ and $\mu=0$, quality factor equations of the HONAF system

The quality factor is generally defined as $Q=\pi f_0\tau$ ($Q=f_0/\Delta f$ can also be defined from the transmission spectrums, $\Delta f$ is the full width at half maximum (FWHM) of the passband) [SF3], and $1/\tau = \sum_{i=1}^{\infty} 1/\tau_i$, where $\tau$ represents the total lifetimes, $\tau_i$ represents the lifetimes of each leaky components, and $f_0$ represents the resonance frequency of cavity [SF1]. Assume that the quality factor of a single isolated cavity is $Q_0$, for the model in this paper (the self-loss of the cavities is ignored), it can be obtained

$$Q_0 = \pi f_0 \tau, \tag{S50}$$

$$1/\tau = 1/\tau_1 + 1/\tau_2 \quad (N=1), \tag{S51}$$

$$1/\tau = 1/\tau_1 + 1/\tau_a = 1/\tau_2 + 1/\tau_a \quad (N \geq 2), \tag{S52}$$

(12) and (17) in the text are also utilized, that is

$$T_{4(N)} = \left| \frac{2^N/\left(\tau_1^{1/2}\tau_2^{1/2}\tau_a^{N-1}\right)\left[j2\pi(f-f_0)+1/\tau_0+1/\tau_1+1/\tau_a\right]^{-1}}{\left[j2\pi(f-f_0)+1/\tau_0+1/\tau_2+1/\tau_a\right]\left[j2\pi(f-f_0)+1/\tau_0+2/\tau_a\right]^{N-2}} \right|, \tag{S53}$$

$$\tau_1 = \tau_2 = \tau_a. \tag{S54}$$

Substituting (S50), (S51) and (S54) into (S45); (S50), (S52) and (S54) into (S53), and then combine and simplify to get

$$T_{4(N)} = \left[\frac{1/(2Q_0)^2}{(f-f_0)^2/f_0^2 + 1/(2Q_0)^2}\right]^N \quad (N \geq 1). \tag{S55}$$

Assume that the above equation is $T_{4(N)}(f=f_1)=1/2$, $f_1$ is the position of the FWHM frequency points from the transmission spectrums, and utilize

$$Q_N = f_0/|2f_1 - 2f_0|. \tag{S56}$$

Simplify to get

$$Q_N = Q_0\left(2^{1/N}-1\right)^{-1/2} \quad (N \geq 1). \tag{S57}$$

The formula is equation of quality factor of the HONAF in the case of cascaded any number of cavities.

When $N=1$ to 5000, the change spectrum of $Q_N$ is shown in Fig. S2.

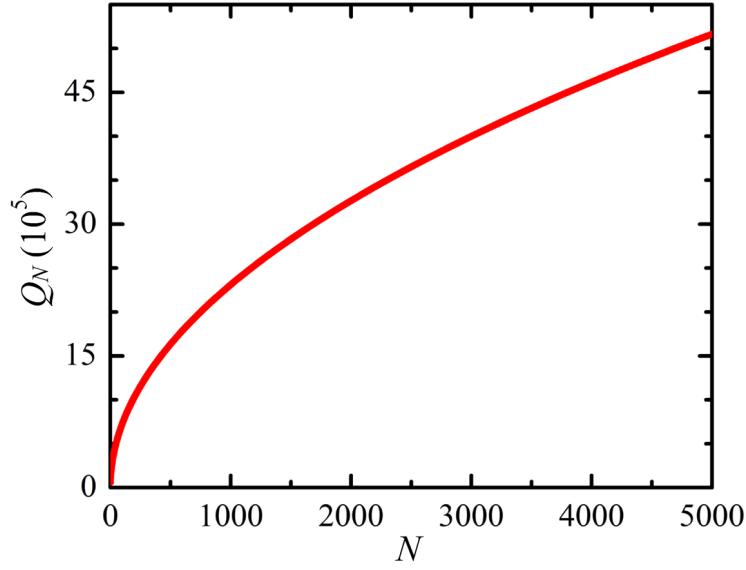

Fig. S2. The $Q_N$ of the HONAF varies with the number of cascaded cavities $N$, assuming $Q_0$=60792.

Obviously, with the increase of $N$, $Q_N$ increases monotonically and the line spectrum is upward convex type. In the case of large $N$ value, the growth of $Q_N$ will become slower, which is consistent with the theoretical analysis.

If the loss of the cavities themselves is considered, then the up-bound of $Q$ can be obtained. Firstly, assume that $\mu \rightarrow 0$, and the loss of the cavities themselves is described by the $Q_s$, so $Q_s = \pi f_0 \tau_0$ [SF1]. Suppose that with the increase of the number of cascaded cavities, the transmission efficiency of the filter finally decreases to 2% due to the existence of loss [SF2], and then $T_{4(N)}(f=f_1) = 1\%$, (S57) becomes

$$Q_N = Q_0 \left(10^{2/N} - 1 - Q_0^2/Q_s^2\right)^{-1/2}. \tag{S58}$$

The first derivative of $Q_N$ from (S58) is

$$Q_N' = \frac{10^{2/N} \ln 10}{N^2} Q_0 \left(10^{2/N} - 1 - Q_0^2/Q_s^2\right)^{-3/2}. \tag{S59}$$

Obviously, $Q_N'$ exists a singularity, which is the zero point $N_0$ of the polynomial of $N$ in parentheses of (S59). Therefore, in the left deleted neighborhood of $N_0$, $Q_N' > 0$, while in the right deleted neighborhood of $N_0$, $Q_N'$ is a positive imaginary value. It means that the $Q_N$ increases monotonically in the left deleted neighborhood of $N_0$, however, the $Q_N$ becomes an imaginary number in the right deleted neighborhood of $N_0$, which has non-physical meaning. To find the singularity, get

$$10^{2/N_0} - 1 = Q_0^2/Q_s^2. \tag{S60}$$

The number of cavities that can be cascaded depends largely on the loss of the cavities.

The larger the loss, the smaller the number of cavities that can be cascaded. Assuming that $Q_s$ is two orders of magnitude larger than $Q_0$, obviously, over a large range of $N$ values, $Q_N$ is increasing. According to the (S60), we can get $N_0 \approx N_{max}=46000$, substitute it into the (S58), so the $Q_{46000} \approx 177418898$. However, in the actual situation, it is far from possible to cascade 46000 cavities, because with the increase of $N$, it is very difficult to achieve $\mu \to 0$. According to our numerical simulation and discussion in the Section VI of the Supplementary Materials, due to the effects of $\mu$, up-bound of the $Q$ should be at the $1 \times 10^6$ level in the filter we designed.

### IV. Transmission spectrums of the HONAF system with $\mu$ playing a leading role

According to (S43), under some parameter values of $N$, $\tau_1$, $\tau_2$, $\tau_a$ and $\mu$, the transmission spectrums of all the ports of the HONAF (Specific expressions refer (S47) to (S49)) is shown in Fig.S3.

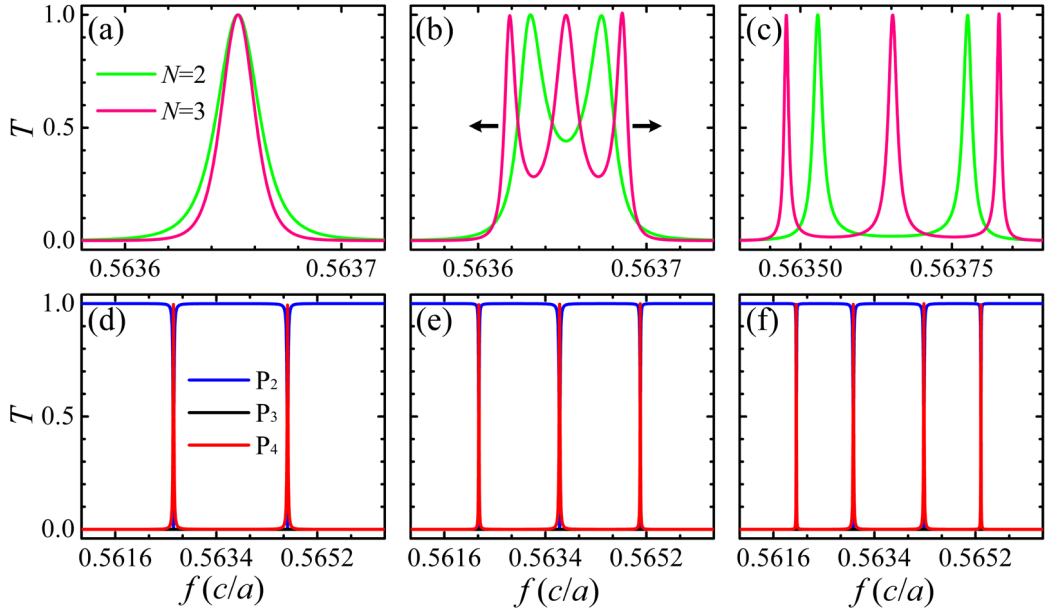

Fig. S3. The transmission spectrums of all the ports of the HONAF. Assuming $f_0=0.56365 \times c/a$, $\tau_1=\tau_2=8.5765 \times 10^{-7}$ s, ignoring the influence of $\tau_0$, (a)-(c) is the transmission spectrums of port $P_4$ under different $\tau_a$ and $\mu$ when $N=2,3$. (d)-(f) respectively represent the transmission spectrums of all the ports when $N=2,3$ and 4, and when $\tau_a$ and $\mu$ are determined. Parameter values (a) $\tau_a=8.5765 \times 10^{-7}$ s, $\mu=0$; (b) $\tau_a=8.5765 \times 10^{-4}$ s, $\mu=3.0670 \times 10^6$ s$^{-1}$; (c) $\tau_a=1.5765 \times 10^{-3}$ s, $\mu=1.6670 \times 10^7$ s$^{-1}$; (d)-(f) $\tau_a=2.8677 \times 10^{-1}$ s, $\mu=1.3674 \times 10^8$ s$^{-1}$.

As can be seen from Fig. S3(a) to S3(c), with the increase of $\mu$, $T_{4(N)}$ is divided from a single formant ($\mu=0$) into multiple formants ($\mu \neq 0$, and the number of formants is equal to $N$). When $\mu$ is diminutive, the formants appear peak spikes and split to both ends, as shown in Fig. S3(b) black arrow. When $\mu$ is large enough, the formants split completely, as shown in Fig. S3(c). By comparing the $N=2$ and $N=3$ transmission spectrums of port $P_4$ in Fig. S3(b) and S3(c), it is found that when $N$ value is large, it is more likely to split into multiple formants. This means that in the actual HONAF structure, even if the coupling distance between the nearest neighbor cavities is large, as $N$ increases, its transmission spectrums will inevitably split from single formant to multiple formants. By comparing the transmission spectrums of $P_4$ port in Figs. S3(c) and S3(d), it is found that when $\mu$ continues to increase ($\tau_a$ changes correspondingly; $\tau_1$ and $\tau_2$ remain unchanged, and $\tau_0=0$), the distance between the formants will increase, which the bandwidth will decrease. It can be seen from Figs. S3(e) and S3(f) that the formants on both sides possess a diminutive bandwidth, while the formants in the middle possess a large bandwidth. Furthermore, the central frequency points of each formants after splitting are symmetric about $f_0$.

Next, analyze the impact of the fluctuation of the coupling strength between the cavities. We first introduce a fluctuation term $\Delta\mu$ to describe the coupling strength between the cascaded cavities fluctuations. In this case, the general coupling strength is $\mu+\Delta\mu$, which is substituted into (S43). According to our analysis of the first- to fourth-order filters in Section II of the text, it is obvious that the fluctuation term will affect the situation of the frequency splitting, that is the resonance frequency points of the filter will change, but the transmission efficiencies are not change.

For more general purposes, the interaction between the $i^{th}$ and the $(i+1)^{th}$ of the cascaded cavities is set as $\mu_i$. Then the parameter $c$ into $c_i = j\mu_i(-j\mu_i+2/\tau_a)$, in this case, we can also use the formulas of Section I deduction, but we can only get recursive formulas

$$H_{n+2} = bH_{n+1} + c_{n+1}H_n, \quad H_1 = a, \quad H_2 = ab + c_1, \tag{S61}$$

$$F_{n+2} = bF_{n+1} + c_n F_n, \quad F_1 = 1, \quad F_2 = a. \tag{S62}$$

Since the coefficient term $c$ is already a function of $n$, we cannot solve the general term and thus cannot obtain the general term formula for $T_{4(N)}$. This makes it impossible for us to directly calculate the transmission efficiencies of the cascaded $N$ cavities, but we can explore the transmission efficiencies when the number of cascades is not large to illustrate the impact of the difference in $\mu_i$. When $N=3$, the transmission efficiency $T_{4(3)}$ is

$$\begin{aligned}
T_{4(3)} = \Big| 2/(\tau_1\tau_2)^{1/2} \left(-j\mu_1 + 2/\tau_a\right)\left(-j\mu_2 + 2/\tau_a\right) \Big/ \Big\{ \big[ j2\pi(f-f_0) \\
+ 1/\tau_0 + 1/\tau_1 + 1/\tau_a \big] \big[ j2\pi(f-f_0) + 1/\tau_0 + 1/\tau_2 + 1/\tau_a \big] \\
\times \big[ j2\pi(f-f_0) + 1/\tau_0 + 2/\tau_a \big] + j\mu_1\left(-j\mu_1 + 2/\tau_a\right) \\
\times \big[ j2\pi(f-f_0) + 1/\tau_0 + 1/\tau_2 + 1/\tau_a \big] + j\mu_2\left(-j\mu_2 \right. \\
\left. + 2/\tau_a\right) \big[ j2\pi(f-f_0) + 1/\tau_0 + 1/\tau_1 + 1/\tau_a \big] \Big\} \Big|^2
\end{aligned} \tag{S63}$$

We obtained transmission spectrum of the $T_{4(3)}$ as shown in Fig.S4, from which it can be found that when the coupling strength between the cascaded cavities is not the same (solid red line), the transmission efficiency will be reduced and the situation of the frequency splitting will also be affected.

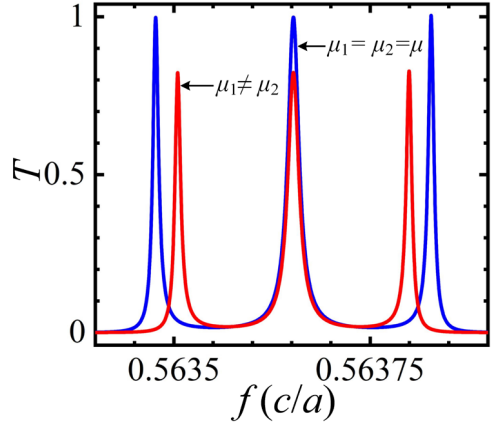

Fig. S4. Transmission spectrum $T_{4(3)}$ of the third-order filter. Parameters $f_0=0.56365\times c/a$, $\tau_1=\tau_2=8.5765\times 10^{-7}$ s, $\tau_a=1.5765\times 10^{-3}$ s, red solid line $\mu_1=1.6670\times 10^7$ s$^{-1}$; $\mu_2=1.0670\times 10^7$ s$^{-1}$; blue solid line $\mu=1.6670\times 10^7$ s$^{-1}$.

From our great deal of calculations, the situation of other high-order filters is similar to that of third-order filters. Therefore, the different coupling strength between cavities will lead to the decrease of transmission efficiency and affect the splitting situation of the resonance frequency.

## V. Transmission efficiency equations of the CROWs system

The CROWs are composed of multiple cavities coupled side by side in a cascade [SF3], [SF4]. Therefore, the two waveguide channels of the above four-port system are deleted, or the waveguide channels are changed into input and output couplers (i.e. the waveguide is the input and output port), which constitute the basic model of the universal CROWs system.

It is assumed that the input and output ends are fully coupled to the cascaded cavities and that the waveguide and each single cavity support only one mode in the same frequency band. The normalized amplitudes of the input and output fields of the waveguide are $S_x$ and $S_y$, respectively. According to the equations above, the output end $S_y$ of the CROWs is

$$S_y = S_{NN} = S_4, \qquad (S64)$$

and $S_x=S_1$, then the transmission efficiency of the output is

$$T = \left|\frac{S_y}{S_x}\right|^2 = \left|\frac{S_4}{S_1}\right|^2. \qquad (S65)$$

Obviously, this equation is consistent with (S43). This equation can be used for the analysis of any CROWs system, is not limited to the input and output is similar the coupler of reciprocity waveguide structure CROWs. Moreover, this equation can be used to analyze the variation of the transmission efficiency of the CROWs with the coupling intensity of the cascade cavities (limited to the weakly coupled system, and the input and output ends are fully coupled to the cascaded cavities).

## VI. Model and performance analysis of the HONAF S

The theoretical model established in this paper can be widely used in various cascaded cavities systems. In order to prove the correctness of the theoretical prediction and the consistency with the simulation results, the basic cascaded unit S (yellow scatterers in Fig. S5(a)) is arbitrarily taken, which constructs the HONAF S, as shown in Fig. S5(b).

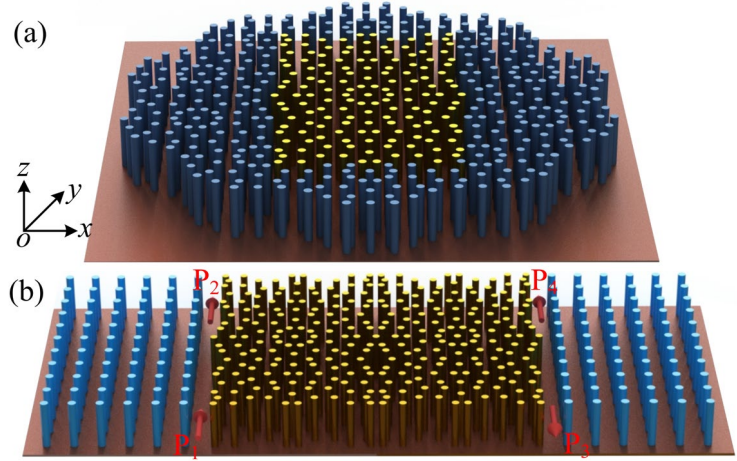

Fig. S5. (a) Decagonal Penrose PQCs structure; (b) 2$^{nd}$-order HONAF model.

The topological waveguide channels are constructed by using GPCs, whose structure parameters and related settings are consistent with the text HONAF.

As can be seen from the structure of HONAF S in Fig. S5(b), similarly, we can obtain the common band gap of the GPCs and the PQCs, the topological edge state and the cavity mode within the band gap by calculating the number of states of the filter. The coupling distance between the nearest-neighbor cavities is enough large, so that $\mu$ is enough diminutive, and it can be considered that $\mu=0$. According to Section II of the text, the formants should be degenerate with different $N$ values, with only one single formant.

The transmission performance of the HONAF S is analyzed below. With initial structure parameters and when order $N=1$ to 6, the transmission spectrums of all the ports of HONAF S are obtained, as shown in Fig. S6.

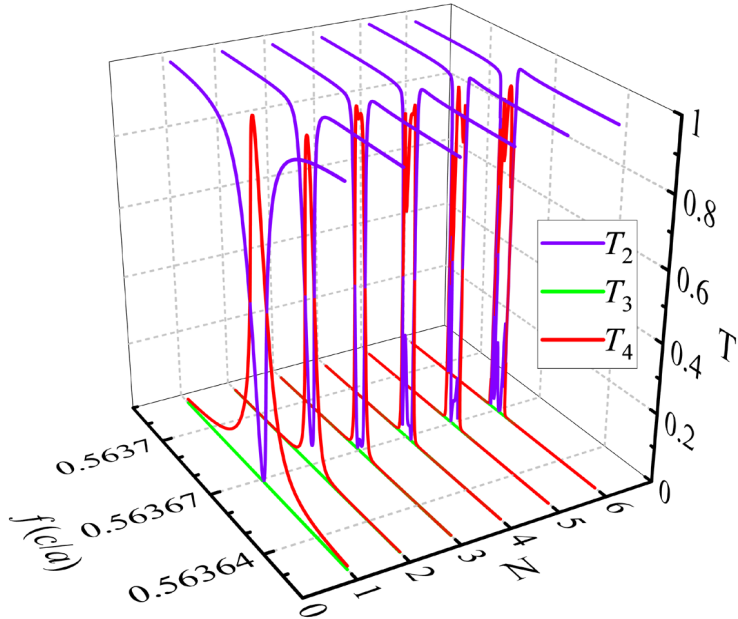

Fig. S6. Transmission spectrums of all the ports of the HONAF S.

The numerical simulation results of the HONAF S in Fig. S6 are basically consistent with the theoretical prediction. When the HONAF S is in resonance state and $N=1$, $T_{2(1)}\approx 0$, $T_{4(1)}\approx 1$, the drop efficiency reaches almost 100%, while $T_{3(1)}\equiv 0$ due to the non-reciprocal waveguide. When $N=2$ to 6, $T_{2(N)}>0$, $T_{4(N)}<1$, the reason is $\tau_1 \neq \tau_a$ ($\tau_1=\tau_2$ is known from the symmetry of the structure) as mentioned in Section II of the text. In addition, when $N$ increases, peak spikes appear in the $P_4$ port transmission spectrum. This is because, with the increase of $N$, the interaction between cascaded cavities is enhanced, so that single formant tends to split into multiple formants. Moreover, with the increase of $N$, the bandwidth of a single formant decreases, which also conforms to the theoretical results.

When $N=1$ to 6, the transmission spectrums of port $P_4$ is shown in Fig. S7.

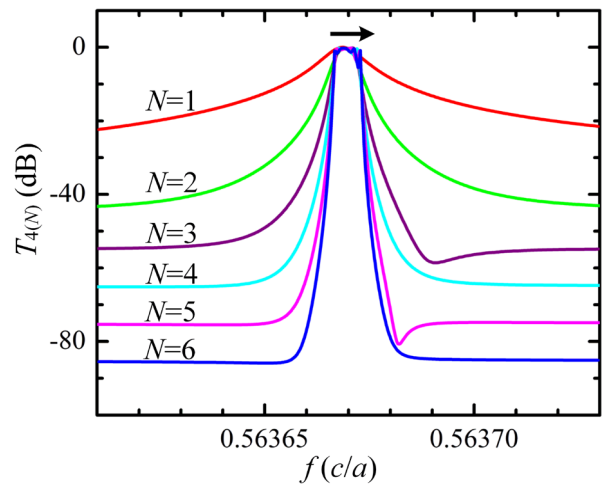

Fig. S7. When $N=1$ to 6, transmission spectrums of port $P_4$ of the HONAF S.

As can be seen from Fig. S7, with the increase of $N$, the out-of-band rejection ratios of the HONAF S also increases. When $N=6$, the out-of-band rejection ratios reaches 85 dB. With the increase of $N$, the formants in the $T_{4(N)}$ transmission spectrums gradually changes from a peak to a rectangular peak (or flat peak), whose position moves toward the high frequency (i.e. blueshifts), as shown by the black arrow in Fig. S7. According to Fig. S6, it is found that the increase of the out-of-band rejection ratios, the appearance of the rectangular peak and the blueshifts of the formants are all caused by the enhancement of the interaction between cascaded cavities with the increase of $N$.

The distribution laws of electric fields $E_z$ in the cascaded cavities resonance of the HONAF S are analyzed below. Through the studies, it is found that when $N$ increases, there are $N$ distribution forms of cascaded cavities electric field, i.e. there are $N$ resonance states of the HONAF S. When the order $N=1$ to 3, the electric fields distribution diagrams of cascaded cavities resonance are obtained, as shown in Fig. S8.

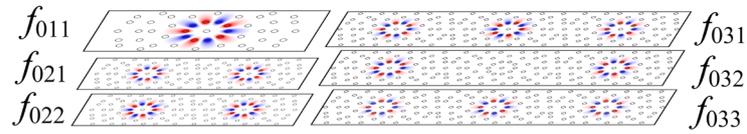

Fig. S8. When $N=1$ to 3, the cascaded cavities of HONAF S is in resonance, the electric fields $E_z$ distribution diagrams. $f_0$ represents the resonance frequency, and the subscript $Ni$ represents the $i^{-th}$ resonance state of the cascaded $N$ cavities, the same as below.

Obviously, the resonance form of cascaded cavities is whispering gallery modes. In this kind of structure, the photons are concentrated in a single cavity and do not remain between adjacent cavities, indicating that the interaction between cascaded cavities is weak. The electric fields in the cascaded cavities are distributed in positive and negative phases, and for a certain value of $N$ ($\geq 2$), the electric fields directions of each pair of adjacent cavities in the cascaded cavities are both symmetric and antisymmetric. Because the interaction between the cascaded cavities is diminutive, the difference between the resonance frequencies of different resonance states under a certain $N$ value is very diminutive. Therefore, from the perspective of the transmission spectrums, each resonance frequency point is degenerate. In fact, as long as $\mu \neq 0$ is satisfied, the frequency splitting situation must exist. With the increase of $N$ value, the number of resonant states of the cascaded cavities increases, and the influence of $\mu$ also increases, so that the frequency splitting trend becomes obvious. As shown in Fig. S7, the transmission spectrums of the HONAF S shows peak spikes.

According to each resonant state of the cascaded cavities, a quality factor $Q$ can be obtained by numerical calculation to measure the filtering performance of this resonant state. Therefore, the quality factor of each resonant state in Fig. S8 is calculated: $Q_{011}=60792$, $Q_{021}=121338$, $Q_{022}=121326$, $Q_{031}=216732$, $Q_{032}=121368$, and $Q_{033}=272423$. However, according to (S54), $Q|_{N=1}=61910$, $Q|_{N=2}=118419$, and $Q|_{N=3}=111840$ can be calculated from the transmission spectrums.

When $N \neq 1$, the $Q$ value obtained by the above two methods describes a completely different state. The former represents the frequency selection performance of a single resonant frequency point in the cascaded cavities, and the corresponding frequency point is completely split. This calculation method is not applicable to models such as the HONAF S (degeneracy of each resonant frequency point). The latter describes the frequency selection performance of cascaded cavities resonance frequency degeneracy, which should be selected in practical application for the models such as the HONAF S. The difference superposition of the resonance frequencies points (blueshifts of the resonance frequencies points, without complete degeneracy) increases the bandwidth of the formants after superposition. Hence, the exception of $Q|_{N=1}<Q|_{N=2}$, in the case of $N≥2$, the actual quality factor $Q$ is going to decrease as $N$ goes up. Therefore, in this case, more cavities through cascades have little contribution to the increase of quality factor of the HONAF S, the number of cascaded cavities $N_{max}=2$. Increasing the coupling distance between the cascaded cavities is an attempt to improve the quality factor (i.e. reducing the interaction factor $\mu$ between the cascaded cavities). However, as mentioned in Section II of the text, excessive coupling distance will lead to a decrease in the coupling rate between the waveguide channels and the cavities and a decrease in drop efficiency. However, for CROWs, the bandwidth increases and the transmission spectrums changes from a peak to a rectangular peak are excellent [SF3].

## VII. Band structures of the gyromagnetic photonic crystal with square lattice and number of states of first-order filter

We use the initial structure parameters to calculate the bands of the GPCs with square lattice in the absence and presence of an external dc magnetic field as show in Fig. S9(a). Obviously, when there is no external magnetic field, the second and third bands are degenerate at point M, and when a dc magnetic field is applied, the degeneracy points of the second and third bands are opened, resulting in a topological band gap. The number of states of the filter can clearly see the common band gap of the GPCs and the PQCs, the topological edge state and the cavity mode based on PQCs located on the band gap. Therefore, we have calculated the number of states using the first-order filter as an example as show in Fig. S9(b) to S9(d) from where the expected information intuitively can be obtained.

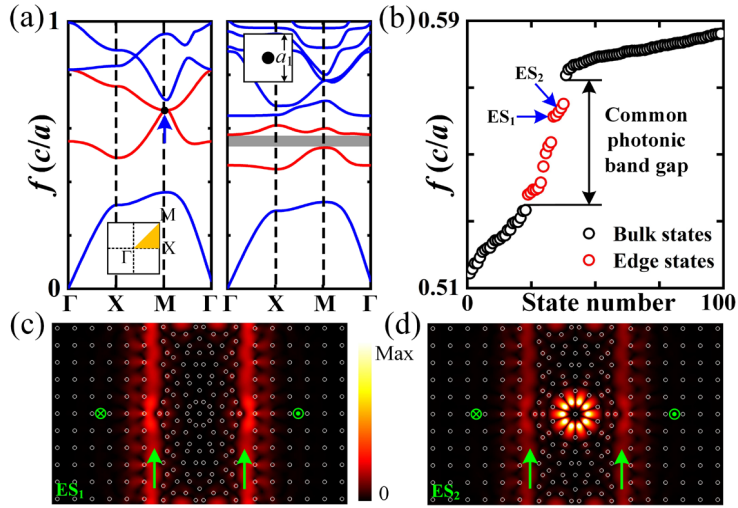

Fig. S9 (a) The band structures of the GPC with square lattice. The left band structure has no applied dc magnetic field, and the second and third bands are degenerate at point M, while the right band structure shows the second and third bands opening a band gap under the applied dc magnetic field. (b) the number of states of first-order filter. Topological edge states and defect modes of the PQCs are located in the band gap, corresponding to the field $E_z$ distributions (c) and (d).

## VIII. Distribution laws of the electric fields $E_z$ in the HONAF

The distribution laws of electric fields $E_z$ in the cascaded cavities resonance of the HONAF are analyzed. When the order $N$=1 to 6, the electric fields distribution diagrams of the cascaded cavities in resonance can be calculated, as shown in Fig.S10.

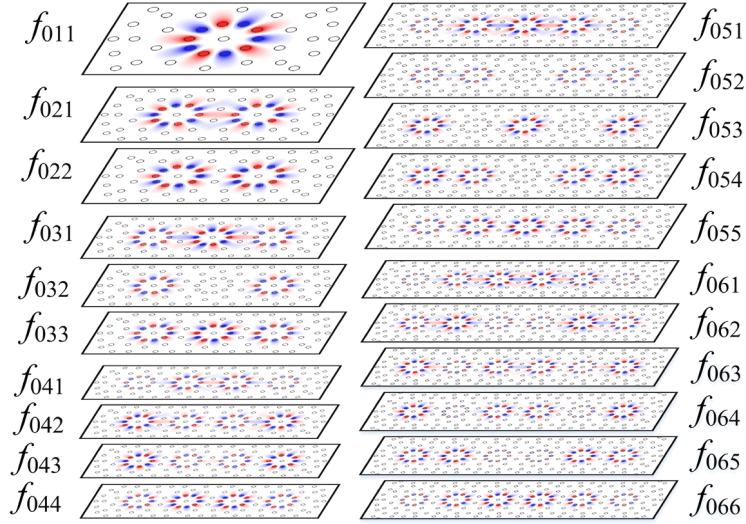

Fig. S10. When $N$=1 to 6, the cascaded cavities of the HONAF is in resonance, the electric fields $E_z$ distribution diagrams.

The resonance mode of cascaded cavities is also whispering gallery modes. The electric fields in the cascaded cavities are distributed in positive and negative phases, and for a certain value of $N$ ($\geq 2$), the electric fields directions of each pair of cavities are both symmetric and antisymmetric. As can be seen from Fig. S10, when the HONAF is in the intermediate resonance state the photons are concentrated in the cavities at the left and right ends, while in the resonance state at the left and right ends, the photons are concentrated in the cavities in the middle, and the trend of this law is more obvious with the increase of $N$. In addition, because the coupling distance between the cascaded cavities is close, the cascaded cavities almost form a whole structure. The photons are not only localized in the cavities, but also localized between adjacent cavities. Therefore, the cascaded cavities can be regarded as a whole cavity structure, which possesses multiple resonance frequencies. The positions of these resonance frequencies are determined by the structure of the whole cavity itself. In this case, the distribution of multiple resonance frequencies in the whole cavity is more complex, which may possess symmetrical properties or be dispersed randomly. The results are consistent with those obtained in Section II of text.

## IX. The simple scheme of experimental design

The simple scheme of experimental design is shown in Fig.S11.

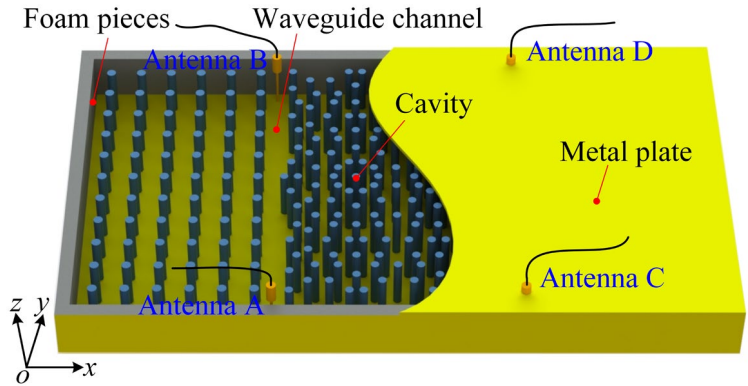

Fig.S11 Simple schematic of experimental realization of the second-order filter. An array (blue cylinder) of the square lattice GPCs and the PQCs, placed between two parallel copper plates (yellow). Parallel copper plates are used to limit electromagnetic waves in the *z* direction. The foam pieces that can absorb microwaves are placed around it as the absorbing boundary. Four dipole antennas are placed on four ports and are connected to the vector network analyzer. Antenna A transmits the excitation source through the vector network analyzer, and antennas B, C and D are used as probes to measure the *S* parameters of each port. An electromagnet is used to generate a dc magnetic field along the *z* direction (not shown).

The lattice constant and the scatterer radius are based on the parameters of the optimal simulation results. For the convenience of fabrication, the GPCs and PQCs scatterers are placed in the air according to the square lattice and their geometric arrangement rules, respectively. The foam pieces that can absorb microwave is placed around the PCs as the absorption boundary to prevent the electromagnetic wave from circulating on the boundary of the PCs and to avoid the system being interfered by the outside environment. The outer plane constraint in the *z* direction is implemented by two parallel copper plates. The scatterers are embedded in the copper plate for fixing. This 3D structure is considered to be quasi-2D and supports TM mode electromagnetic waves consistent with those in PCs. It supports the TEM mode in which the electric field points to the *z* direction of the outer plane and the magnetic field is parallel to the *x* and *y* directions. This polarization mode is the same as the TM mode in 2D PCs. The electromagnetic field of the TEM mode is also uniform along the *z* direction between the two plates, just as it is in a 2D system. This 3D structure is therefore very close to a 2D system and is considered quasi-2D [SF5], [SF6]. An electromagnet is used to generate a dc magnetic field in the z direction, which is applied to the GPCs. Dipole antennas are placed on four ports and connected to the vector network analyzer to measure S parameters. The four dipole antennas pass through two parallel copper plates in the *z* direction through the upper plate, and are fixed on the copper plate. Special attention should be paid to the fact that the size of all the scatterers of the PQCs should be as consistent as possible, and they should be placed on the plate according to the geometric arrangement rules of the PQCs, so that the Q of all the cavities constructed by the PQCs is consistent with the resonance frequency.

# References

[SF1] J. D. Joannopoulos, S. G. Johnson, J. N. Winn, and R. D. Meade, *Photonic crystals: molding the flow of light*, Princeton University Press, New Jersey, USA, 2008.

[SF2] J. Zhou, W. Han, Y. Meng, H. Song, D. Mu, and W. Park, "Optical properties of direct and indirect coupling of cascaded cavities in resonator-waveguide systems," *J. Lightw. Technol.*, **32**, 4104 (2014).

[SF3] F. Morichetti, C. Ferrari, A. Canciamilla, and A. Melloni1, "The first decade of coupled resonator optical waveguides: bringing slow light to applications," *Laser Photonics Rev.*, **6**, 74 (2012).

[SF4] H. Liu and S. Zhu, "Coupled magnetic resonator optical waveguides," *Laser Photonics Rev.*, **7**, 882 (2013).

[SF5] F. Li, H. Wang, Z. Xiong, Q. Lou, P. Chen, R. Wu, Y. Poo, J. Jiang, and S. John, "Topological light-trapping on a dislocation," *Nat. commun.*, **9**, 2462 (2018).

[SF6] Z. Wang, Y. Chong, J. D. Joannopoulos, and M. Soljačić, "Observation of unidirectional backscattering-immune topological electromagnetic states," *Nature*, **461**, 772 (2009).